%% file: ms.tex
\documentclass{article} 
\usepackage{iclr2023_conference,times}

\input{math_commands.tex}

\usepackage{hyperref}
\usepackage{url}
\usepackage[pdftex]{graphicx}
\usepackage{tcolorbox}
\usepackage{marvosym}
\usepackage[]{authblk} 
\newcommand*{\affaddr}[1]{#1} 
\newcommand*{\affmark}[1][*]{\textsuperscript{#1}}
\newcommand*{\email}[1]{\texttt{#1}}
\usepackage{hyperref}

\title{Active Bird2Vec: Towards End-to-End Bird Sound Monitoring with Transformers}

\author{%
Lukas Rauch \Letter \affmark[1], Raphael Schwinger\affmark[2],Moritz Wirth\affmark[1,3], Bernhard Sick\affmark[1], Sven Tomforde\affmark[2], and Christoph Scholz \affmark[1,3] \\
\affaddr{\affmark[1]IES, University of Kassel}\\
\email{\small\{lukas.rauch, moritz.wirth, b.sick, c.scholz\}@uni-kassel.de}\\
\vspace{0.1cm}
\affaddr{\affmark[2]INS, University of Kiel}\\
\email{\small\{rsc,st\}@informatik.uni-kiel.de}\\
\vspace{0.1cm}
\affaddr{\affmark[3]IEE, Fraunhofer-Institut}\\
\email{\small\{moritz.wirth, christoph.scholz\}@iee.fraunhofer.de}
}

\iclrfinalcopy 
\begin{document}

\maketitle

\begin{abstract}
We propose a shift towards end-to-end learning in bird sound monitoring by combining self-supervised (SSL) and deep active learning (DAL). Leveraging transformer models, we aim to bypass traditional spectrogram conversions, enabling direct raw audio processing. \textsc{Active Bird2Vec} is set to generate high-quality bird sound representations through SSL, potentially accelerating the assessment of environmental changes and decision-making processes for wind farms. Additionally, we seek to utilize the wide variety of bird vocalizations through DAL, reducing the reliance on extensively labeled datasets by human experts. We plan to curate a comprehensive set of tasks through Huggingface Datasets, enhancing future comparability and reproducibility of bioacoustic research. A comparative analysis between various transformer models will be conducted to evaluate their proficiency in bird sound recognition tasks. We aim to accelerate the progression of avian bioacoustic research and contribute to more effective conservation strategies.


\end{abstract}

\section{Introduction}
Avian vitality trends serve as indicators of baseline environmental health \citep{kahl2021}. Especially changes in bird populations can signal shifts in biodiversity as they play crucial roles in seed dispersal, controlling pests, and pollination, directly impacting an ecosystem's health \citep{sekercioglu2016}. Additionally, as wind farms are emerging to combat climate change, the potential risks to bird species necessitate strategic planning. Therefore, bird monitoring becomes critical in assessing biodiversity and guiding turbine placement to minimize collisions. Traditionally, bird monitoring centers on audio recordings, requiring human experts (i.e., ornithologists) to perform exhaustive field studies. 

Recent strides in deep learning (DL) offer promising results in bird sound recognition \citep{kahl2021}. These DL advancements in bioacoustics primarily operate on spectrograms (e.g., mel-scale spectrograms) of audio recordings and employ convolutional neural networks \citep{stowell2021}. However, converting raw audio data into mel-scale spectrograms constitutes a processing step that requires manual parameter adjustments. For instance, the number of mel filters determines the mel-scale resolution, notably impacting the audio signal's representation. This process complicates the comparability among various methodologies and can degrade the audio signal quality \citep{gazneli2022}. Guided by advancements in natural language processing (NLP), transformer models such as Wav2Vec2 \citep{baevski2020} that utilize self-supervised learning (SSL) on large volumes of unlabeled speech data signify a shift towards end-to-end (e2e) learning approaches directly applied to raw waveform audio \citep{liu2022}. Despite the wealth of (weakly-labeled)\footnote{Weakly-Labeled recordings are of varied lengths for which a label is given, but without a specific time stamp.} bird vocalizations available on Xeno-Canto \citep{vellinga2015}, an online community-driven platform, current DL techniques fail to generalize well enough to detect rare species or unique call signs in challenging environments \citep{ghani2023}. Leveraging transformer models and self-supervised learning (SSL) at scale could potentially excel in capturing high-quality bird sound representations and enhance generalization performance. While SSL can reduce the need for labeled data to achieve high model performance, the vast diversity of bird species still necessitates ornithologists to provide labels to further fine-tune a model on downstream tasks. Deep active learning (DAL) addresses this challenge by enabling DL models to selectively query annotations for instances expected to yield the highest performance gains. In practical settings, high-quality feature representations from SSL are crucial for rapidly adapting models to downstream tasks with DAL \citep{rauch2023}. 

\section{Research Approaches}
\begin{figure*}[!h]
\centering
\includegraphics[width=0.99\columnwidth]{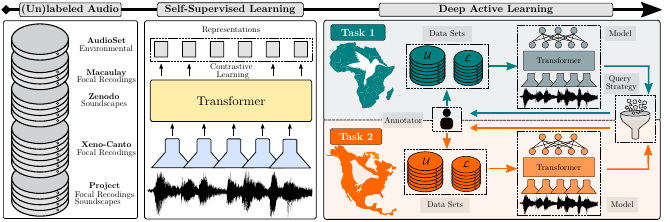}
\caption{A schematic illustration of the proposed \textsc{Active Bird2Vec} approach.}
\label{fig:ga}
\end{figure*}

With \textsc{Active Bird2Vec} (cf. Figure \ref{fig:ga}), our objective is to create high-quality sound representations through SSL and swiftly adapt models for downstream tasks using DAL. Our process advances in three stages. First, we collect and pre-process bird vocalizations from sources such as Xeno-Canto comprising focal and soundscape recordings. While focal recordings capture isolated bird sounds using directional microphones, soundscape recordings employ an omnidirectional microphone to gather ambient sounds \citep{kahl2021}. This initial stage establishes a streamlined evaluation protocol fostering a diverse set of tasks and laying the groundwork for future research. Second, we pre-train a transformer model (e.g., Wav2Vev2) using SSL (e.g., contrastive learning), producing high-quality bird sound representations from raw audio. In this phase, we intend to provide a foundation model that can be adapted to various tasks. Lastly, by leveraging DAL, the pre-trained model actively select the instances it expects to yield the highest performance gains for a specific task. This adaptive approach enables model customization and improved label efficiency \citep{tran2022} for specific regions, call types, or bird species.

We design \textsc{Active Bird2Vec} as an e2e solution, streamlining the monitoring pipeline. Thereby, we eliminate the need for labor-intensive manual feature engineering, aiming to reduce labeling effort and expedite the model's adaptation to novel deployment scenarios in bird sound recognition. In the following, we put forward the key strategies for expanding the scope of this research in greater detail. These key strategies set a roadmap for addressing future challenges. 

\subsection{Unified Evaluation Protocol and Benchmark Suite}
The field of bird sound monitoring lacks a unified evaluation protocol and benchmark results, impeding comparisons across studies. While a wealth of open-source data is available on platforms such as Xeno-Canto, data set standardization is absent, undermining reproducibility and comparability. Currently, researchers must manually curate training and testing datasets from various sources, creating a substantial accessibility barrier for newcomers entering this field. Especially the distinction between focal recordings in training and soundscape recordings in testing poses an additional challenge to ensure the applicability of models in practice. Despite attempts at pipeline automation \citep{burmeister2021}, the field lacks a standardized evaluation protocol and an established benchmark for different tasks. Moreover, the closed-source nature of certain evaluation data sets in current literature \citep{denton2021, kahl2021} and the vast diversity of bird species further exacerbate these challenges. We argue that these challenges impede the progress in bird sound monitoring. 

\begin{tcolorbox}
    \textbf{Our approach:} Addressing these challenges, we propose a bird sound monitoring benchmark suite with a unified evaluation protocol and a diverse set of tasks, following \citet{wang2019} and \citet{rauch2023}. This involves curating various data sources and types, enabling the assessment of the generalization capabilities of DL models when recognizing bird sounds in real-world scenarios. Additionally, we aim to establish baseline results for our benchmark through an extensive empirical study. This serves as a reference point for ongoing work and a means to identify future challenges. By integrating our curated data set collection into Huggingface Datasets \citep{wolf-etal-2020-transformers,lhoest-etal-2021-datasets}, we strive to ease accessibility, refine the evaluation process and foster more consistent experimental designs in forthcoming research. 
\end{tcolorbox}

\subsection{Evaluating End-to-End Transformers}
Bird sound recognition primarily relies on manually converting sound files to images through spectrogram conversions and employing CNNs as a standard practice for image classification. However, current research shows the potential of e2e transformer models in environmental sound tasks \citep{gazneli2022,lopez-meyer2021,elliott2021} and human speech recognition \citep{latif2023,baevski2020,baevski2022,liu2022a}. Despite promising advancements, the application of these models to bird sound recognition remains largely unexplored. For instance, \citep{bravosanchez2021} utilize the SincNet \citep{ravanelli2019} convolutional architecture, previously applied to speech processing, to learn directly from raw waveforms for bird sound monitoring. Additional research indicates that a self-supervised learning (SSL) model trained on speech signals could hold valuable representations for other sound types \citep{liu2022}. This serves as a promising foundation for further exploration in bird sound monitoring, focusing on transformer models with attention mechanisms for raw waveforms. 

\begin{tcolorbox}
    \textbf{Our approach:} Through a comprehensive empirical analysis, we aim to unravel the strengths and shortcomings of e2e transformer models. Our objective is to guide the selection of the most suitable model for specific tasks within our benchmark suite and evaluation protocol. This approach may enrich current insights and create pathways for more refined and optimized applications of transformer models in bioacoustic research.
\end{tcolorbox}

\subsection{Representation Learning with Self-Supervised Learning using Transformers}
With the advancements in SSL of transformer models in NLP, representation learning in the audio domain has gained increasing attention \citep{yang2021,wang2022,baevski2022}. Although prevalent in the field of human speech recognition \citep{baevski2020,latif2023}, the application of SSL with transformer models to bird sound monitoring is mainly unexplored despite a large volume of data from passive acoustic monitoring~\citep{perez‐granados2021a}. SSL techniques may provide a unique learning signal that differs from traditional supervised learning, capturing more diverse data properties \citep{balestriero2023}. A foundation model with high-quality bird sound representations could substantially streamline the fine-tuning process, allowing quicker model convergence on challenging downstream tasks and reducing the need for labeled training data. 
\begin{tcolorbox}
    \textbf{Our approach:} We propose to rigorously investigate the application of SSL with transformer models on raw audio data with the objective of learning high-quality bird sound representations in our proposed benchmark suite. Guided by the principles underpinning successful speech recognition models such as Wav2Vec2, our research is set to harness extensive unlabeled datasets. Moreover, we seek to publish a foundation model \texttt{Bird2Vec} as a versatile tool that can be employed for multiple downstream tasks, propelling advancements in this research domain.
\end{tcolorbox}

\subsection{Model Adaptability through Deep Active Learning}
While SSL may reduce the required number of labeled samples with high-quality representations, the pre-trained model has to adapt quickly to a diverse set of downstream tasks (e.g., geographic locations, call types, bird species) in an application scenario. The heterogeneity of bird species, sound types (e.g., song and call), and recording types (e.g., focal and soundscape recording) complicate model adaption, especially in unique application settings such as monitoring potential wind farm locations. This follows the trend towards reliable transformer models, even when pre-trained on a large scale \citep{tran2022}. While SSL offers a promising avenue to generate high-quality representations that reduce the need for labeled samples, we still need to obtain time-consuming and expensive labels from human experts in challenging environments. Thus, applying DAL to the SSL-generated representations could enrich the learning signal for DAL, potentially bolstering the model performance, reducing the labeling cost, and speeding up the adaption process even further in practice \citep{tran2022, rauch2023}. 

\begin{tcolorbox}
    \textbf{Our approach:} We plan to evaluate the effectiveness of DAL for bird sound recognition, focusing on augmenting model adaptability and curtailing labeling expenses. Recognizing the potential benefits, our primary interest pivots around the interplay between DAL and high-quality representations derived from SSL, thereby offering a more efficient and flexible adaptation to diverse real-world tasks. Through this integrative approach, anchored by our proposed benchmark suite, we aim to pave the way for more efficient, adaptable, and cost-effective methods in bioacoustics.
\end{tcolorbox}

\section{Conclusion}
In light of environmental conservation efforts, efficient bird sound monitoring has become crucial. In this paper, we introduced the \textsc{Active Bird2Vec} approach and laid out a roadmap with the key strategies to promote our approach. We design \textsc{Active Bird2Vec} to advance bird monitoring in bioacoustics by combining self-supervised learning and deep active learning, promoting the generation of high-quality bird sound representations and reducing labeling expenses. By proposing an end-to-end solution, we aspire to streamline the monitoring pipeline and eliminate the dependency on labor-intensive manual feature engineering. Furthermore, we seek to foster reproducibility and comparability across studies by establishing a standardized evaluation framework and benchmark suite that also serves as the foundation of our approach. 

\section*{Acknowledgements}
This research has been funded by the German Ministry for the Environment, Nature Conservation, Nuclear Safety, and Consumer Protection through the project "DeepBirdDetect - Automatic Bird Detection of Endangered Species Using Deep Neural Networks" (67KI31040C).
\bibliography{literatur}
\bibliographystyle{iclr2023_conference}


\end{document}

%% file: math_commands.tex
\usepackage{amsmath,amsfonts,bm}

\def\eqref#1{equation~\ref{#1}}

\def\1{\bm{1}}

\DeclareMathAlphabet{\mathsfit}{\encodingdefault}{\sfdefault}{m}{sl}
\SetMathAlphabet{\mathsfit}{bold}{\encodingdefault}{\sfdefault}{bx}{n}

